\documentclass[12pt]{article}
\usepackage{psfig}
\usepackage{verbatim}
\usepackage{color}
\usepackage{rotate}
\usepackage{graphics}
\usepackage{epsf}
\usepackage{epsfig}
\usepackage{graphicx}
\usepackage{latexsym}
\usepackage{amscd}
\usepackage{amssymb}
\newcommand{\beq}{\begin{equation}}
\newcommand{\eeq}{\end{equation}}
\newcommand{\beqn}{\begin{eqnarray}}
\newcommand{\eeqn}{\end{eqnarray}}
\newcommand\noi{\noindent} 
\newcommand\la{\langle}
\newcommand\ra{\rangle}
\newcommand\eps\varepsilon

\newcommand\imag{{\rm i}}

\def\GeV{\,\mbox{GeV}}

\def\lsim{\mathrel{\rlap{\lower4pt\hbox{\hskip1pt$\sim$}}
    \raise1pt\hbox{$<$}}}         
\def\gsim{\mathrel{\rlap{\lower4pt\hbox{\hskip1pt$\sim$}}
    \raise1pt\hbox{$>$}}}         

\begin{document}

\hfill {LA-UR-02-2041}

\vspace*{2cm}
\begin{center}
{\Large
\bf 
Parton model versus color dipole formulation\\

\smallskip
of the Drell-Yan process}
\end{center}
\vspace{.5cm}

\begin{center}
 {\large
J\"org Raufeisen\footnote{\tt email: jorgr@lanl.gov}$^{,a}$,
Jen-Chieh Peng$^{a,b}$, and
Gouranga C. Nayak$^{a}$}\\
\medskip

$^a${\sl Los Alamos National Laboratory,
Los Alamos, New Mexico 87545, USA}\\
$^b${\sl Department of Physics, University of Illinois, Urbana, 
Illinois 61801, USA}
\end{center}

\vspace{.5cm}

\begin{abstract}
\noi
In the kinematical region where the center of mass energy is 
much larger than all other scales, 
the Drell-Yan process can be formulated 
in the target rest frame in
terms of the same color dipole cross section as low Bjorken-$x$ 
deep inelastic scattering. 
Since 
the mechanisms for heavy dilepton production appear very different
in the dipole approach and in the conventional parton model, 
one may wonder
whether these two formulations really represent the same physics. 
We perform a comparison
of numerical calculations in the color dipole approach with calculations
in the next-to-leading order parton model.
For proton-proton scattering, the results are very similar
at low $x_2$ from fixed target to RHIC energies, confirming
the close connection between these two very different approaches. 
We also compare the transverse momentum distributions of Drell-Yan 
dileptons predicted in both formulations.
The range of applicability of the dipole formulation 
and the impact of future Drell-Yan data from RHIC for determining the 
color dipole cross section 
are discussed. A detailed derivation of the
dipole formulation of the Drell-Yan process is also included.    
\medskip

\noi
\end{abstract}

\clearpage

\section{Introduction}\label{sec:intro}

With the advent of RHIC, the Drell-Yan(DY) process \cite{dy} can be studied in a new kinematical 
regime, the so-called Regge regime, where the dilepton mass $M$ is small
compared to the {\em cm} energy $\sqrt{s}$, but still much larger 
than $\Lambda_{QCD}$. 
The DY process at RHIC and LHC energies
is therefore of similar interest as DIS at HERA, where 
one can study $\gamma^*$-proton scattering in the Regge regime.
The new experimental possibilities motivate further theoretical investigations
of the DY process. First of all, one needs a framework
to calculate nuclear shadowing in the DY process, the onset of which 
can already be
observed at fixed-target energy \cite{e772pa}. This is especially important in view
of the RHIC heavy ion program. 
The color dipole formulation of the DY process introduced in \cite{boris,bhq} is
suitable to address this issue and a considerable amount of work
in this direction has been performed \cite{kst1}-\cite{krtj}.
In addition, the low-$x_2$ DY cross sections are sensitive to
integrals over the 
color dipole cross section which are not accessible in DIS \cite{bhq}. 
Therefore, future
DY data can be used to further constrain this quantity.
Most importantly, while the dipole approach has already been used in 
sophisticated analyses like the extraction of energy loss from E866/NuSea 
Drell-Yan data \cite{eloss1}, 
its validity for proton-proton interactions has never been 
established. This will be remedied with the present work.

In this paper we compare 
next-to-leading order (NLO) parton model calculations for DY 
dilepton production in proton-proton 
($pp$) and proton-deuteron ($pd$) collisions with calculations in the 
dipole approach over a wide energy range. Although the two approaches
are believed to be equivalent in a certain kinematical range, the underlying  mechanisms
appear to be quite different, and there is no known way to prove this equivalence analytically. 
However, both approaches are supposed to describe the same
process, so they should yield similar numerical results\footnote{Differences 
between parton model and dipole approach are seen in the angular 
distribution of DY pairs, if saturation effects are included 
in the dipole cross section \cite{krt3}. When we talk about {\em equivalence}
of parton model and dipole approach, we mean equivalence up to higher twist effects,
which are neglected in the parton model, but (at least partially) included
in the phenomenological parameterization of the dipole cross section we employ.
}.

Before we compare the results of the numerical calculations, the key features of
the two approaches are briefly summarized.
The well known mechanism for continuum dilepton production, which was
first found more 
than thirty years ago by 
Drell and Yan \cite{dy}, was formulated in a frame where both colliding
hadrons are fast moving (infinite momentum frame). According to 
Feynman's picture of high energy collisions, the colliding objects can 
be viewed as collections of noninteracting partons with negligibly small transverse
momenta. To lowest order, DY dileptons are produced by quark-antiquark 
annihilation, and the cross section reads,
\beq\label{eq:dylo}
\frac{d^2\sigma_{DY}}{dM^2\, dx_F}=\frac{4\pi\alpha^2_{em}}{9M^2s}\,\frac{1}{x_1+x_2}
\sum_{f=1}^{N_f}Z_f^2\left[q_f(x_1)\bar q_f(x_2)+q_f(x_2)\bar q_f(x_1)\right].
\eeq
The distribution function of a quark (antiquark) 
of flavor $f$ of the target or the projectile is denoted by $q_f$ ($\bar q_f$), $N_f$ is
the number of active flavors and $Z_f$ is the quark charge. The longitudinal 
momentum fractions of the projectile (target) parton, $x_1$ ($x_2$),  
can be expressed in terms of Lorentz invariant scalar products as
\beq\label{eq:x1x2}
x_1=\frac{2P_2\cdot q}{s}\quad;\quad
x_2=\frac{2P_1\cdot q}{s}, 
\eeq
where $P_1^\mu$ ($P_2^\mu$) is the projectile (target) four momentum, $q^\mu$ 
is the four momentum of the dilepton, $q^2=M^2>0$, and $x_F$ is the Feynman-$x$,
$x_F=x_1-x_2$.

For most qualitative descriptions, it is sufficient to consider the DY process in
terms of the lowest order annihilation process, Eq.~(\ref{eq:dylo}). Calculations with
Eq.~(\ref{eq:dylo}), however, underestimate measured DY cross sections by an 
overall factor. 
It is necessary to employ  
the NLO framework for the DY process, in order to make quantitative predictions,
see \cite{mmp} for a review.
In addition, the DY cross section differential in the dileptons transverse momentum
receives huge corrections from higher order processes. Indeed, to lowest order, one 
would not expect dileptons with large transverse momentum $q_\perp$, 
in contrast to what is observed in experiment. Even though the occurrence of perturbatively
large transverse momenta can be explained in NLO, it is not straightforward 
to calculate the shape of the $q_\perp$-distribution in the parton model. 
A resummation of large logarithms
in $q_\perp/M$ \cite{resumcol} or alternatively the introduction of an intrinsic 
transverse momentum \cite{app}
is necessary to avoid the divergence of the differential cross section at 
$q_\perp=0$.

In the parton model, all nonperturbative effects are parameterized in the parton distribution
functions $q_f$, $\bar q_f$, which evolve according 
to the DGLAP evolution equations.
For DY in nuclear collisions, the parton distribution functions of the proton
are simply replaced by empirical nuclear parton distribution functions \cite{ekrt}.
This approach does not explain the dynamical origin of the 
nuclear effects

\begin{figure}[ht]
  \centerline{\scalebox{0.8}{\includegraphics{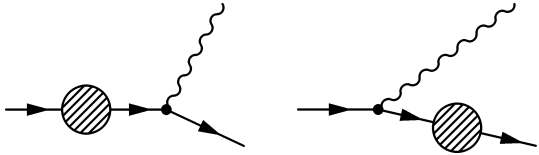}}}
    \center{\parbox[b]{13cm}{\caption{
      \label{fig:dy}\em
      In the target rest frame,
      DY dilepton production looks like bremsstrahlung. A quark 
      or an antiquark from the
      projectile hadron scatters off the target color field 
	(denoted by the shaded circles)
	and radiates a
      massive photon, which subsequently decays into the lepton pair.
	The photon decay is not shown.
	The photon can be radiated before or after the quark (antiquark) scatters.}  
    }  }
\end{figure}

Nuclear effects,
effects from higher orders in perturbation theory, as well as other
possible nonperturbative effects, 
are more readily treated when the Drell-Yan process is viewed in the target rest frame.
Note that
although cross sections are Lorentz invariant, the partonic interpretation of
high energy scattering processes does depend on the reference frame. In the
rest frame of the target, the production
mechanism for high mass continuum dileptons looks like bremsstrahlung \cite{boris,bhq}, 
see Fig.~\ref{fig:dy}. In the high energy limit,
when one can neglect terms that are suppressed by a factor $1/$energy, each of the
two graphs factorizes into a production vertex for the virtual photon times an amplitude
for scattering a quark off the target. These scattering amplitudes combine in the squared matrix
element in exactly the same way as in DIS, which makes it possible to express the DY cross
section in terms of the same cross section $\sigma^N_{q\bar q}$ 
for scattering a $q\bar q$-dipole off a nucleon ($N$) as in low-$x_{Bj}$ DIS,
\beq\label{eq:dylctotal}
\frac{d\sigma(qN\to\gamma^*X)}{d\ln\alpha} =
\int d^2\rho\,\left|\Psi_{\gamma^*q}(\alpha,\rho)\right|^2\,
\sigma^N_{q\bar q}(\alpha\rho,x)\ .
\label{eq:10}
\eeq 
Here, $\alpha$ is the light-cone momentum fraction the virtual photon takes away from
its parent quark, and $\rho$ is the transverse separation between $\gamma^*$ and final quark.
The electromagnetic radiation, $q\to\gamma^*q$, is described by the light-cone wavefunction 
$\Psi_{\gamma^*q}(\alpha,\rho)$, see Eqs.~(\ref{dylct}) -- (\ref{eq:lcwfsum}),
which can be calculated perturbatively. 
Summation over photon polarizations is understood in Eq.~(\ref{eq:dylctotal}).
The dipole cross section $\sigma^N_{q\bar q}$ is of nonperturbative origin and has
to be taken from phenomenology. 
The energy scale $x$ of the dipole cross section will 
be discussed in the next section. A detailed derivation of Eq.~(\ref{eq:dylctotal}) 
is given in the appendix. 

Using a phenomenological 
parameterization for the dipole cross section in Eq.~(\ref{eq:dylctotal}) is a very economical 
way to account for higher order and nonperturbative effects. 
The dipole approach can even be applied at low values of $M$ where
perturbative QCD is not valid \cite{prompt}.
It was found in a recent analysis
\cite{krt3} that most of
E772 DY data (except some points at low $M$) 
are reasonably well described in the dipole approach
without introducing an arbitrary overall normalization factor. In addition it was found that
the transverse momentum distribution does not diverge at $q_\perp=0$, even without
intrinsic transverse momentum. 

We emphasize that the dipole approach does not describe an additional production mechanism
for heavy dileptons. Rather, the two approaches are believed to describe the same
physics in different reference frames.
Therefore, calculations in the NLO 
parton model and in the dipole approach should give similar results for the DY cross
section. This is what we numerically check in this paper.
In the following section, we compare numerical calculations of the DY cross section
(integrated over the transverse momentum of the dilepton)
in both approaches.
In section 3, we also compare the predictions 
of dipole approach and parton model for the DY transverse momentum distribution
at RHIC.

\section{Numerical comparison of the two approaches}

In order to perform calculations that can be compared with experimental
data, one has to embed the partonic cross section, Eq.~(\ref{eq:dylctotal}), 
into the hadronic environment. In the infinite momentum frame, the 
momentum fraction of the projectile quark is $x_1$, see Eq.~(\ref{eq:x1x2}).
However, when the scalar product defining $x_1$ is evaluated in the 
target rest frame, one finds $x_1=\alpha z$, where $z=x_1/\alpha$ is the momentum 
fraction of the incoming proton carried by the projectile quark.  
The different 
meanings of $x_1$ in the target rest frame and in the infinite momentum frame
is a manifestation of the frame dependence of partonic mechanisms.
In the target rest frame, $x_1$ is the momentum fraction that the lepton pair 
takes from the projectile proton.   
Thus, one obtains for the proton-nucleon DY cross section
\beqn\label{eq:dylctotalhadr}\nonumber
\frac{d^2\sigma(pN\to l^+l^-X)}{dM^2dx_F} &=&
\frac{\alpha_{em}}{3\pi M^2}\frac{x_1}{x_1+x_2}
\int_{x_1}^1\frac{d\alpha}{\alpha^2}\sum\limits_{f=1}^{N_f}
Z_f^2\,\left[q_f\left(\frac{x_1}{\alpha},\widetilde Q\right)
+\bar q_f\left(\frac{x_1}{\alpha},\widetilde Q\right)\right]\\
&\times &
\int d^2\rho\,\left|\Psi_{\gamma^*q}(\alpha,\rho)\right|^2\,
\sigma^N_{q\bar q}(\alpha\rho,x)\ .
\eeqn 

We still need to know the scale $\widetilde Q$ at which the projectile parton
distributions are probed and the energy $x$ at which the dipole cross 
section enters. These quantities are not known exactly, instead 
we have to rely on plausible arguments to determine their values. In order to find 
$\widetilde Q$, note that the transverse distances $\rho$ 
that contribute to the DY cross section are controlled by the extension 
parameter
\beq
\label{eq:eta}
\eta^2=(1-\alpha)M^2+\alpha^2m_f^2.
\eeq
The numerically dominant term in the LC wavefunctions, 
Eqs.~(\ref{dylct}, \ref{dylcl}), is the one that contains the Bessel function
K$_1(\eta\rho)$. Since this function decays exponentially at large arguments,
the largest distances that can numerically contribute are of order 
$\sim 1/\eta$. For fluctuations with $\alpha\to 1$, these distances can become
of the order of a typical hadronic radius, in analogy to the aligned jet 
configurations in DIS \cite{aljet}. On the other hand, the minimal value of $\alpha$ is 
$x_1$, so that the largest virtuality entering the calculation is 
$\widetilde Q^2=\eta^2_{\rm max}=(1-x_1)M^2$. We choose this quantity to
be the hard scale at which the projectile parton distribution is probed.
The parton distribution functions (PDFs) are taken from CERNLIB \cite{cernlib}.
The quark mass is set to $m_f=0$ 
in all our calculations, see \cite{krt3} for its 
numerical influence.

For the quark density of the projectile, 
we employ the leading order 
parameterization that corresponds to the NLO parameterization used
in the parton model calculation. 
This means {\em e.g.} we use CTEQ5L
in the dipole approach when comparing it with a NLO parton model calculation
using CTEQ5M. One should 
use leading order PDFs in the dipole approach, because 
they are scheme independent and have a probabilistic interpretation.

The energy scale $x$ of the dipole cross section in 
Eq.~(\ref{eq:dylctotalhadr}) is determined from the analogy to DIS. In
DIS, the argument of the dipole cross section is 
$x_{Bj}=Q^2/W^2$, where $Q$ is the virtuality of the photon and $W$ is the 
$\gamma^*$-proton {\em cm} energy. Therefore, we choose 
$x=M^2/\hat s=\alpha x_2$, where
$\hat s=sx_1/\alpha$ is the quark-proton {\em cm} energy squared.

Note that in the previous analysis \cite{krt3}, $M^2$ and $x_2$,  
instead of $\widetilde Q^2$ and $x$, were used. 
The different choice of scales in this
paper has the effect of increasing the cross section 
by a factor of up to $2$ for dilepton mass $M\sim 4$ GeV. 
This is mostly due to the different choice of $\widetilde Q^2$.
Using $\alpha x_2$ instead of $x_2$ is only a $\sim 10\%$ effect
at $x_2<0.1$. These uncertainties vanish at larger masses, $M\sim 8$ GeV.

\begin{figure}[ht]
  \centerline{\scalebox{0.8}{\includegraphics{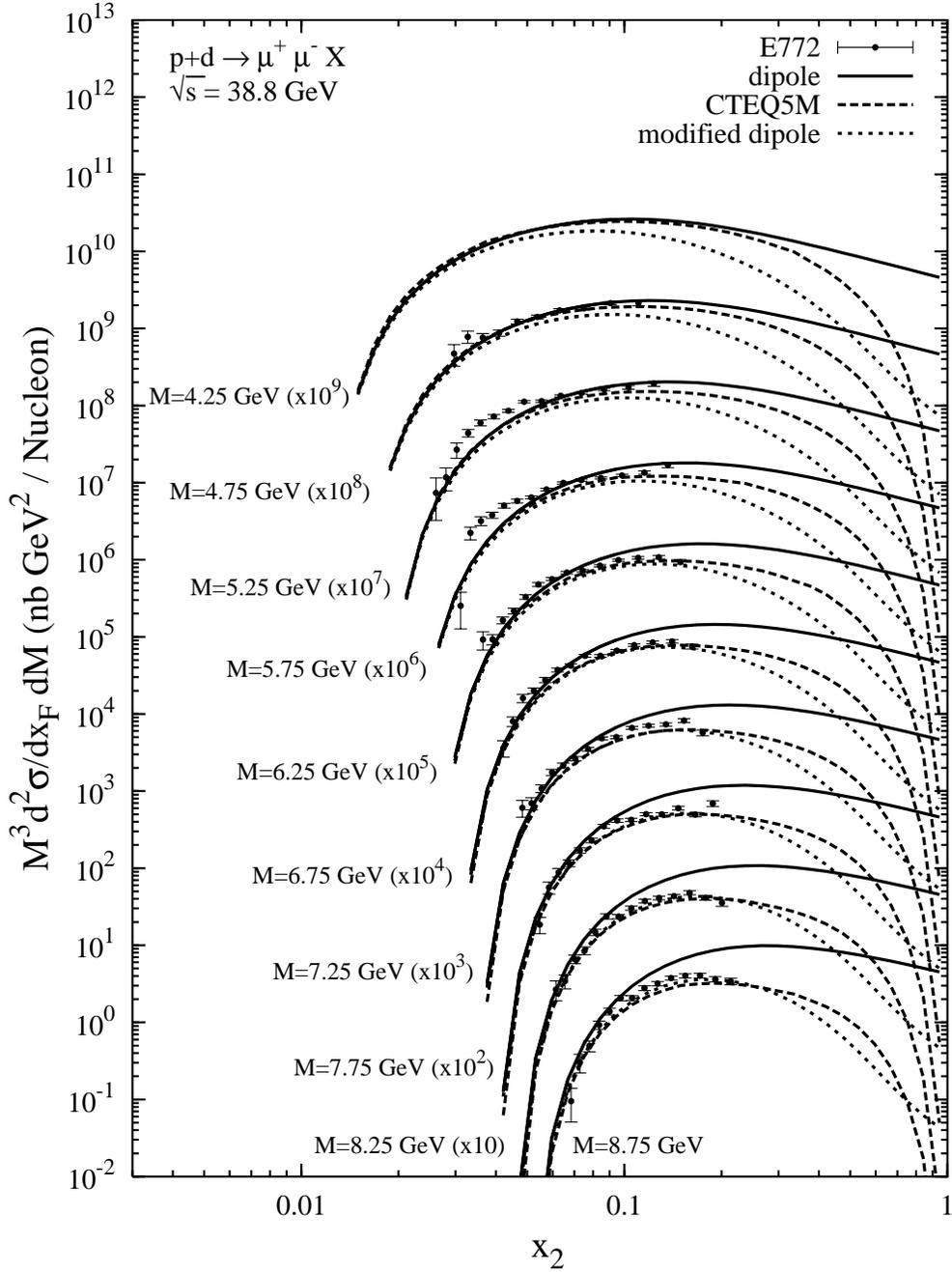}}}
    \center{\parbox[b]{13cm}{\caption{
      \label{fig:e772}\em Calculations in the dipole approach to DY and its 
modification Eq.~(\ref{eq:modified}) compared to NLO parton model results
at fixed target energy ($\sqrt{s}=38.8$ GeV). 
The CTEQ5M parameterization \cite{cteq} is used for the parton model 
calculation. The data are from E772 \cite{e772}.
The curves and data for the different mass bins have been rescaled for
better visibility.
In all calculations, none of the free parameters of the dipole approach were
adjusted to DY data. Only DIS data have been used.
	}  
    }  }
\end{figure}

\clearpage

\begin{figure}[ht]
  \centerline{\scalebox{0.8}{\includegraphics{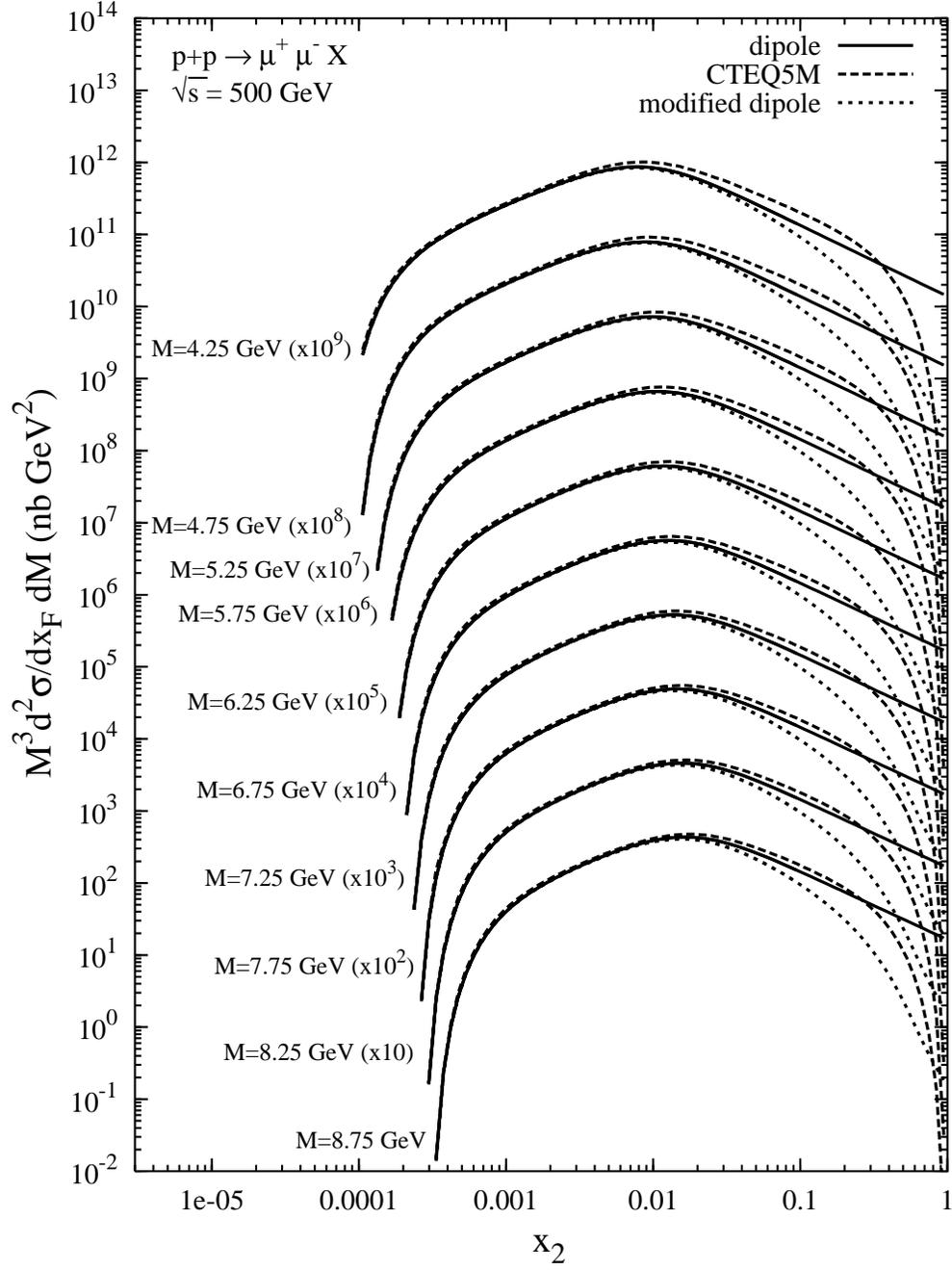}}}
    \center{\parbox[b]{13cm}{\caption{
      \label{fig:rhic}\em The same as Fig.~\ref{fig:e772}, 
but now at RHIC energy
($\sqrt{s}=500$ GeV), where much lower values of $x_2$ can be reached.
}  
    }  }
\end{figure}

\begin{figure}[ht]
  \centerline{\scalebox{0.8}{\includegraphics{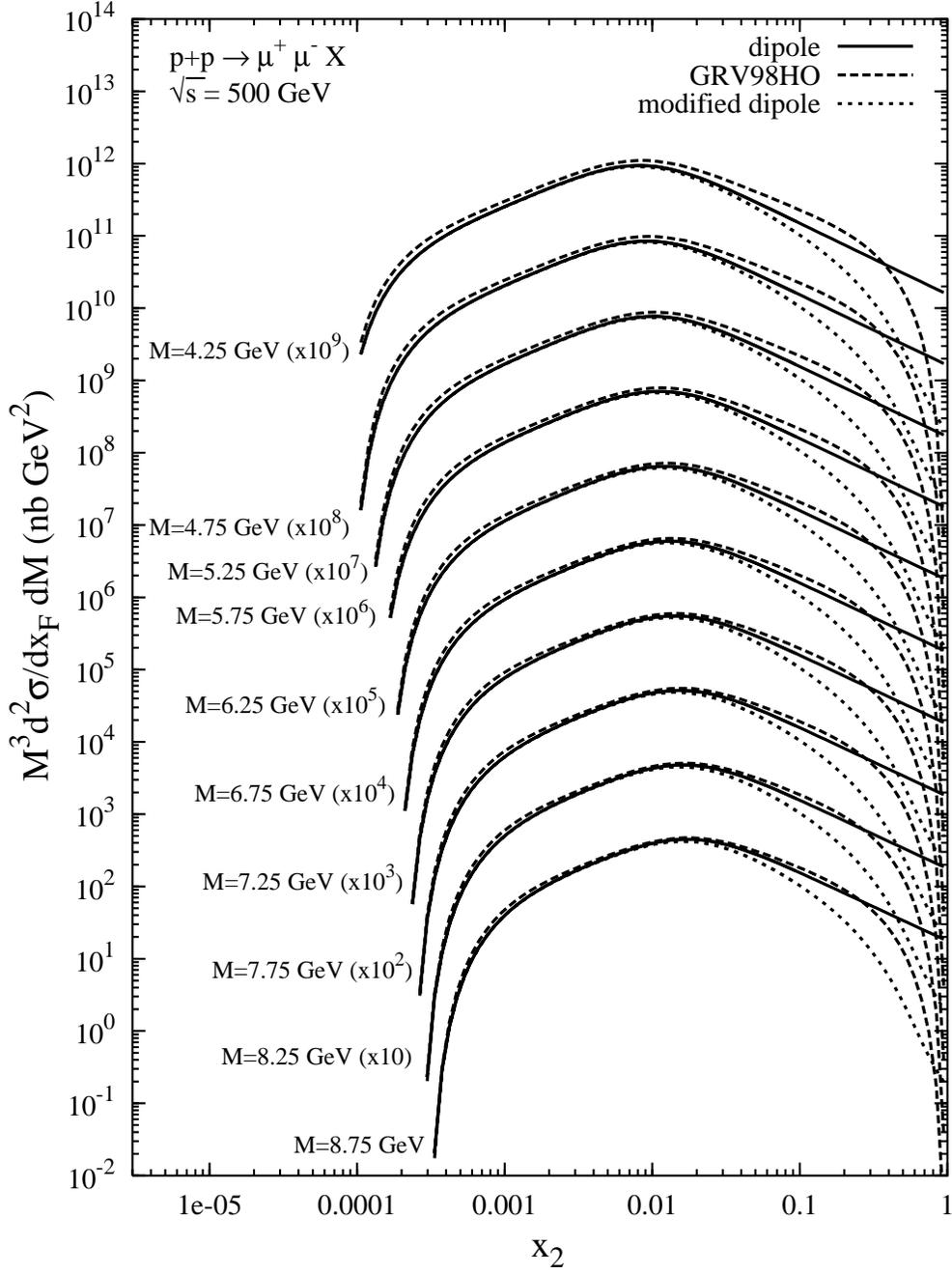}}}
    \center{\parbox[b]{13cm}{\caption{
      \label{fig:rhicgrv}\em The same as Fig.~\ref{fig:rhic}, 
but with GRV PDFs \cite{grv}. For the dipole approach, we employed
the LO PDFs GRV98LO for the projectile parton distribution.  
}  
    }  }
\end{figure}

For the calculations in the dipole approach shown in 
Figs.~\ref{fig:e772} -- \ref{fig:rhicgrv}, we employ the parameterization
of the color dipole cross section
by Golec-Biernat and W\"usthoff \cite{Wuesthoff1},
\beq
\label{eq:gbw}
\sigma^N_{q\bar q}(\rho,x)=
\sigma_0\left[1-\exp\left(-\frac{\rho^2Q_s^2(x)}{4}\right)\right],
\eeq
with $\sigma_0=23.03$ mb. This parameterization rises quadratically
at small separations $\rho$ as demanded by color transparency \cite{ct} 
and saturates at large separations. The saturation scale that controls
the flattening of the dipole cross section is given by
\beq
\label{eq:qs}
Q_s^2(x)=1\GeV^2\left(\frac{0.0003}{x}\right)^{0.288}.
\eeq
We point out that no DY data have been used to determine the parameters in 
Eq.~(\ref{eq:gbw}). Only DIS data from HERA were fitted to extract the dipole cross
section.

Note that the parameterization Eq.~(\ref{eq:gbw}) contains only the
Pomeron part of the dipole cross section.  
As a consequence, the dipole approach predicts
identical cross sections for particle and antiparticle induced DY.
This is, of course, reproduced in the parton model at low $x_2$.
In addition to this Pomeron part, there is also a contribution
from Reggeon exchange, corresponding to valence quarks in the target 
\cite{krtj}.
In principle, one could fit a parameterization of this
Reggeon part to the data, but the predictive power of such a procedure
would be very low, since the Reggeon part, in contrast to the Pomeron
part, depends on the colliding hadrons. We therefore do not attempt 
to determine the Reggeon contribution to the dipole cross section in 
this paper. 

The NLO parton model calculation was performed with the CTEQ code 
\cite{Handbook} provided earlier to the E866/Nusea collaboration
by W.K.~Tung \cite{dbarubar}. We use the dilepton mass $M$ 
as hard scale in the strong coupling constant and in the parton distributions.
This is the standard choice. The code assumes $\overline{MS}$ renormalization 
scheme.

Fig.~\ref{fig:e772} shows calculations at fixed target energy in the dipole 
approach and in the NLO parton model, as well as experimental DY data 
from E772 \cite{e772}. 
The lowest value of $x_2$ that can be reached is about 0.02. 
Dipole and parton model calculations are very similar at low $x_2$, where 
the dipole approach is supposed to be valid, but strongly deviate as 
$x_2\to 1$. Clearly, we cannot apply the dipole approach at large $x_2$,
since it does not include valence quark contributions to the DY cross section
(see \cite{krt3} for a more detailed discussion of effects that are taken 
into account by the dipole approach). Note also that the parameterization
Eq.~(\ref{eq:gbw}) was fitted only to DIS data below $x_{Bj}<0.01$, hence 
the curves shown in Fig.~\ref{fig:e772} 
are already an extrapolation of the fit.

In Fig.~\ref{fig:e772} there are several E772 data points 
at low $x_2$ that exceed the calculations in both the 
parton model and dipole approaches. 
This excess was also visible for some points in \cite{krt3}.  
Indeed, experiment E772 was not designed
to measure the absolute normalization of the DY cross section, but to 
measure nuclear effects. New data from E866/NuSea will not suffer from such
problems. A comparison between preliminary E866/NuSea data and the NLO parton
model in \cite{e866} shows good agreement.

The disagreement between the two approaches sets in around $x_2\sim 0.1$
in Fig.~\ref{fig:e772}. 
Somewhat surprisingly, the dipole approach yields values that exceed 
the prediction from the parton model (at large $x_2$), 
even though the dipole approach does not include several
contributions to the DY cross section. This can be understood,
if one observes that the dipole cross section is related to the target gluon
density $G$. At small separations $\rho$ one has \cite{fs} 
\beq
\label{eq:gdens}
\sigma^N_{q\bar q}(\rho,x)=\frac{\pi^2}{3}
\alpha_s\left(\frac{\lambda}{\rho^2}\right)\rho^2
xG\left(x,\frac{\lambda}{\rho^2}\right).
\eeq
Eqs.~(\ref{eq:gbw}) and (\ref{eq:qs}) 
contain only the part of the gluon density that increases
at low $x$. At large $x$, however, the gluon density decreases like 
$G(x)\propto (1-x)^5$ \cite{west}. 
At high virtuality, the power is even higher than 5, say
$\sim 8$.
In order to estimate the uncertainty originating from this effect, we 
follow \cite{Cronin} and replace 
the saturation scale Eq.~(\ref{eq:qs}) by
\beq
\label{eq:modified}
Q_s^2(x)\to Q_s^2(x)(1-x)^5.
\eeq
We refer to this recipe as {\em modified dipole approach} and the 
numerical results are shown by the short dashed curves in Figs.~\ref{fig:e772}
-- \ref{fig:rhicgrv}. The replacement Eq.~(\ref{eq:modified}) has  
virtually no influence at $x<0.01$, where the parameters in Eq.~(\ref{eq:gbw})
have been fitted in \cite{Wuesthoff1}, but at $x=0.1$, it reduces the cross 
section by almost a factor of 2. In fact, in \cite{ducati} calculations
in the dipole approach were performed employing a dipole cross section
calculated from a gluon density that contains saturation effects. 
For $\sqrt{s}=38.8\GeV$, the results of \cite{ducati} lie below the ones
obtained in \cite{krt3} with the saturation model of Golec-Biernat 
and W\"usthoff \cite{Wuesthoff1} by a factor of $\sim 2$.
We believe this disagreement 
is largely due to the decreasing gluon density at $x_2>0.01$,
{\em i.e.} due to uncertainties in the dipole cross section.

Apparently, the low-$x_2$ range accessible at fixed target energies is very
limited, and therefore the dipole formulation is bestowed with
several uncertainties. Unfortunately, there are no DY data at really low $x_2$.
Even CDF data \cite{cdf} are at about the same values of $x_2$ as the E772 data,
because they are 
mostly in the $Z$-boson mass range.
At RHIC, however, much lower values of $x_2$ will 
be reached, where the dipole approach can readily be applied. In 
Fig.~\ref{fig:rhic}, we show predictions from the dipole approach and the 
NLO parton model (using CTEQ5M PDFs) 
for RHIC energy. Calculations with the modified dipole
cross section, Eq.~(\ref{eq:modified}), are also shown. 
The disagreement between dipole approach and parton model sets in
around $x_2\sim 0.01$, {\em i.e.} around $x_F\sim 0$.
At low $x_2$ all three
curves are in good agreement, 
confirming the similarity between the two approaches.

In order to show the uncertainty arising from the choice of PDFs, we did the 
same calculation, employing GRV98HO \cite{grv}, see Fig.~\ref{fig:rhicgrv}.
In this case, the differences between the two approaches are slightly larger
than for CTEQ5M. With GRV98HO, the NLO parton model calculation exceeds the
dipole approach by at most a factor of $1.5$ for $x_2<0.01$; for CTEQ5M,
the corresponding value is only about $1.2$. The third standard
PDF we employed was MRST(c-g) \cite{mrst}. The results are comparable
to GRV98HO and are not shown
here, because they do not provide any further insight. 
For most values of $x_2$ and $M$, which do not lie at the very edge of
the phase space, the NLO calculations are only $\sim 10\%$ -- $20\%$
higher than the dipole approach.
In summary,
we find that the uncertainty from the choice of PDFs is $\sim 25\%$
at the low values of $x_2$ shown in Figs.~\ref{fig:rhic} and \ref{fig:rhicgrv}.

\clearpage

We also point out that for these low values of $x_2$,
the DY $K$-factor is typically $1.4$ to $1.5$ and can become even
smaller than unity for very low $x_2$. 
In the parton model, the
$K$-factor is defined as 
the DY cross section calculated in NLO divided by the same quantity
calculated in LO. 
The $K$-factor originates (mostly) 
from the analytic continuation
from the spacelike (DIS) to the timelike (DY)
region of $q^2$ in the annihilation and virtual
corrections at NLO. The QCD Compton process, which dominates at low $x_2$
because of the large target gluon density, 
does (almost) not contribute to the $K$-factor.
Therefore, in the kinematical region of interest,
the $K$-factor is considerably below
its usually assumed value of $2$ to $3$.

At this point, we stress that the term ``$K$-factor'' 
is defined only in the parton model. In the dipole approach,
there is no freedom to adjust the overall normalization 
of the DY cross section as proposed in \cite{ducati},
because higher order corrections are contained in the 
parameterization of the dipole cross section. 

Most uncertainties in the dipole approach arise from
uncertainties in this parameterization. 
It remains to be seen whether future low $x_2$
DY data from RHIC can be accurately described
with a (probably improved) phenomenological 
parameterization of the dipole cross section.
We believe that future DY data can serve as 
an important source of knowledge about the color dipole cross section.

\section{The DY transverse momentum distribution}

The DY cross section differential in the dilepton transverse momentum $q_\perp$
can provide even more detailed information about the shape of the dipole
cross section than the $q_\perp$-integrated cross section. In the dipole 
approach, the $q_\perp$-differential cross section reads \cite{kst1},
\beqn\nonumber\label{eq:dylcdiff}
\frac{d^3\sigma(qp\to \gamma^*X)}{d\ln\alpha\, d^2q_\perp}
&=&\frac{1}{(2\pi)^2}
\int d^2\rho_1d^2\rho_2\, 
{\rm e}^{{\rm i}\vec q_\perp\cdot(\vec\rho_1-\vec\rho_2)}
\Psi^{*}_{\gamma^* q}(\alpha,\vec\rho_1)
\Psi_{\gamma^* q}(\alpha,\vec\rho_2)\\
&\times&
\frac{1}{2}
\left\{\sigma_{q\bar q}(\alpha\rho_1,x)
+\sigma_{q\bar q}(\alpha\rho_2,x)
-\sigma_{q\bar q}(\alpha(\vec\rho_1-\vec\rho_2),x)\right\},
\eeqn
where $q_\perp$ is the $\gamma^*$ transverse momentum {\em wrt} the
direction of the projectile quark.
This formula was first published in \cite{kst1}. We give a detailed 
derivation in the appendix.
The partonic cross section, Eq.~(\ref{eq:dylcdiff}), 
has to be embedded into the hadronic environment
in the same way as in Eq.~(\ref{eq:dylctotalhadr}).
Again, the sum over photon polarizations is understood to be contained 
in the light-cone wavefunctions, see Eq.~(\ref{eq:lcwfsum}).

We would like to compare the transverse momentum distribution calculated
in the dipole approach with the same quantity computed in the parton model.
Since the CTEQ code used in the previous section is not capable of calculating
the DY transverse momentum distribution, we developed our own code based on
the formulas given in \cite{app}. Again, we use the dilepton mass $M$ as hard 
scale.
It was already mentioned in the introduction, 
that there are no large transverse momenta in the leading order parton
model. The order $\alpha_s$ correction yields a transverse momentum 
distribution that diverges at $q_\perp\to 0$ and has the wrong shape, when
compared to data \cite{app}. This can be remedied by resumming 
contributions from soft gluon radiation \cite{resumcol},
which yields a good description of the data.

In this paper, however, we follow \cite{app} and
apply a more phenomenological recipe. We introduce
a soft, nonperturbative, primordial transverse momentum distribution of the 
partons in the colliding protons, which we parameterize by a Gaussian,
\beq
f(k_{\perp}^2)=\frac{1}{4 \pi \sigma_q^2} 
{\rm e}^{\frac{-k_{\perp}^2}{4 \sigma_q^2}}.
\eeq
The mean transverse primordial momentum squared
of a single constituent in the proton
is then 
\beq
\la k_{\perp}^2\ra_{\rm primordial}=2\sigma_q^2\quad.
\eeq
The perturbatively calculated $q_\perp$-distribution is then smeared out
by this primordial momentum and the regularized
dilepton cross section is given by
\beqn\nonumber\label{eq:smear}
\left(\frac{d^4 {\sigma^{DY}}}{dM^2 dx_F d^2q_{\perp}}\right)_{\rm reg} &=& \int d^2 p_{\perp}
\left(\frac{d^4 {\sigma^{DY}}}{dM^2 dx_F d^2p_{\perp}}\right)_{\rm A+C} 
[f(({\vec q_{\perp}}-{\vec p_{\perp}})^2)-f(q_{\perp}^2)]\\
&+& f(q_{\perp}^2) \left(\frac{d^2 {\sigma^{DY}}}{dM^2 dx_F }\right)_{\rm NLO+LO}. 
\eeqn
The subscript $A+C$ refers to ``annihilation+Compton'' correction. The virtual
corrections do not contribute to the first term in Eq.~(\ref{eq:smear}).
With this recipe, the perturbative result is reproduced for large transverse momenta while
the divergence at $q_\perp=0$ is removed. The first term on the {\em rhs} of Eq.~(\ref{eq:smear})
vanished after integration over $q_\perp$ and 
$\int d^2q_\perp f(q_\perp)$ is normalized to unity, so that 
the total DY cross section is correctly reproduced.

\begin{figure}[t]
  \centerline{\scalebox{0.8}{\includegraphics{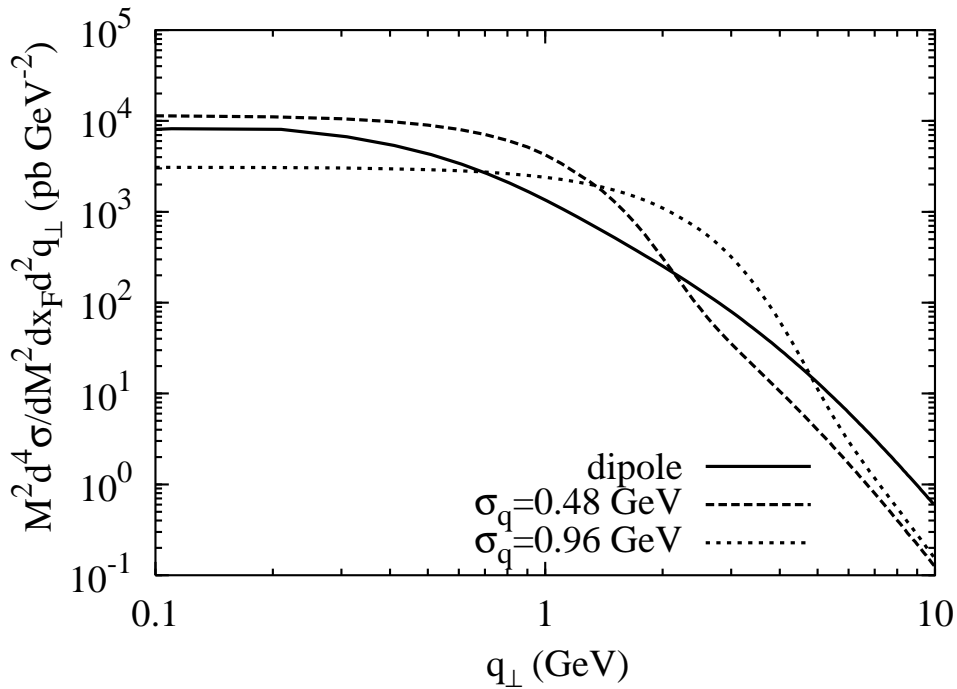}}}
    \center{\parbox[b]{13cm}{\caption{
      \label{fig:perp}\em
	Comparison of the DY transverse momentum distribution calculated
	in the dipole approach and in the NLO parton model. The parton model
	calculations have been performed for two different values of the
	quark's intrinsic momentum. The values of the kinematical variables
	for all curves in this figure are $\sqrt{s}=500$ GeV, $x_F=0$ and
	$M=4.25$ GeV.}  
    }  }
\end{figure}

The results of our calculations in the dipole approach and in the parton
model are shown in Fig.~\ref{fig:perp}. We performed the parton model
calculation for two different values of the smearing parameter $\sigma_q$.
The lower value, $\sigma_q=0.48$ GeV, was used in \cite{antreasyan} to describe
ISR data at $\sqrt{s}=44$ GeV and $62$ GeV. The parameter $\sigma_q$ is 
presumably independent of energy $s$. Nevertheless, we also calculated for 
twice the ISR value of $\sigma_q$, 
because at the low values of $x$ reached at RHIC, quarks may
have a larger intrinsic transverse momentum. Of course, the
choice of $\sigma_q$ has the largest influence at small $q_\perp$, while 
at very large transverse momentum, the results become independent of 
this parameter. Note however, that $\sigma_q$ still has a rather large
impact on the numerical results at intermediate $q_\perp\sim 2$ -- $3$ GeV.

No intrinsic transverse momentum is included in the dipole approach, even
though one could introduce one. As already pointed out in \cite{krt3}, the 
cross section calculated in the dipole approach does not diverge at 
$q_\perp\to 0$ because of the flattening of the dipole cross section,
see Eq.~(\ref{eq:gbw}). At $q_\perp=10$ GeV, dipole approach and parton 
model differ by about a factor of 5. Employing the modified dipole approach,
Eq.~(\ref{eq:modified}),
brings the curve down by a factor of $0.9$ at $q_\perp=10$ GeV and has no
effect at small $q_\perp$. Therefore, we do not show the corresponding
curves.
Note that at high transverse 
momentum, the numerical value of the DY cross section is completely
insensitive to the large $\rho$ behavior of the dipole cross section. 
Indeed, it was found in \cite{krt3} that an expansion
of Eq.~(\ref{eq:gbw}) to order $\rho^2$ virtually yields the same numerical 
results at large $q_\perp$ when employed in Eq.~(\ref{eq:dylcdiff}).
On the other hand, the behavior at $q_\perp\to 0$ is largely determined by
the large $\rho$ behavior of the dipole cross section.
Therefore, future data on the DY $q_\perp$-distribution can be important for
further constraining the dipole cross section, provided statistics 
will be high enough. 

\section{Summary}

In this paper, we presented a detailed comparison between calculations of
the Drell-Yan cross section in the color dipole approach with calculations
in the conventional NLO parton model. We also compared both approaches
with experimental data when available. The dipole approach and the
parton model are believed to represent the same physics, but viewed from 
different reference frames. The purpose of this paper was to demonstrate
the close connection between the two approaches numerically.

First, we compared calculations for the DY cross section integrated over the 
dilepton's transverse momentum $q_\perp$ and found that at low $x_2$, where
the dipole approach is supposed to be valid, calculations agree well.
Since both approaches rely on different nonperturbative inputs, numerical results
will not agree exactly. There are several uncertainties from free
parameters that can affect the 
numerical value of the Drell-Yan cross section. In the dipole approach,
these uncertainties are:
\begin{itemize}
\item{Different possible choices 
of the hard scale $\widetilde Q$, at which the parton
distribution of the projecetile is probed, can lead to an uncertainty as
large as a factor of $2$ for $M\sim 4\GeV$, even though the dependence 
on $\widetilde Q$ is only logarithmic. This uncertainty vanishes with 
increasing dilepton mass $M$. For $M=8.75\GeV$ it is only $\sim 10\%$.
Note that a larger value of $\widetilde Q$ leads to a smaller value of 
the DY cross section, because the projectile PDF is probed at large
momentum fraction.}
\item{For the low values of $x_2$ we are interested in, different 
parameterizations of PDFs lead to variations of the numerical results
of up to $\sim 25\%$. This uncertainty is also present in the parton model.}
\item{The uncertainty arising from possible choices of the energy scale $x$
is $\sim 10\%$. Since the dipole cross section increases with decreasing
$x$, a lower value of $x$ pulls the DY cross section up.}
\item{The dipole approach allows one to introduce a quark mass $m_f$.
In this work we set $m_q=0$. A quark mass $m_q=200$ MeV would reduce
the numerical value of the DY cross section by 
less than $10\%$ \cite{krt3}.}
\end{itemize}
The main uncertainty, which is difficult to quantify, arises 
from the parameterization of the 
dipole cross section. 
Future DY data from RHIC will certainly help to 
better constrain this quantity.

We found that the disagreement between the two approaches is typically of order 20\%
for $x_2<0.01$. We believe that the parameterization of the dipole cross section 
can be adjusted so that future RHIC data can be described without introducing
an arbitrary overall normalization factor for which we cannot find a theoretical
justification.  
For larger values of $x_2$, the applicability of the dipole approach
is questionable, because this approach neglects several contribution to the DY cross
section which might already be important for $0.01\le x_2 \le 0.1$. 

Note that several data points from E772 \cite{e772} 
at low $x_2$ are not well described 
by neither of the approaches. 
Preliminary E866/NuSea data \cite{e866}, however,
which were measured in the same
kinematical region, are well described by the NLO parton model. Therefore,
the 
agreement between dipole approach and NLO parton model at low $x_2$ gives
confidence into the applicability of the dipole approach. However,
we stress that in order to strictly prove the validity of the dipole approach,
one has to reproduce experimental data and not the parton model calculation.

Naturally, the differences between the two approaches are larger when one considers
the DY transverse momentum distribution. At low $q_\perp$, the numerical result
in the parton model depends quite strongly on the amount of primordial transverse
momentum which is included. This uncertainty is however no longer present at very large
transverse momenta, $q_\perp\sim 10\GeV$, where the two calculations differ by a factor 
of 5 (at $\sqrt{s}=500\GeV$) . It will be seen, if RHIC data can distinguish between 
the two approaches or if the dipole cross section and/or the parton distribution 
function can both be adjusted to fit the data. We point out that both approaches
predict the same asymptotic behavior for the partonic DY cross section, namely
$d\sigma/d^2q_\perp\propto 1/q_\perp^4$ for large transverse momentum \cite{kst1}.

\bigskip
{\bf Acknowledgments:}
We are indebted to Fred Cooper, Mikkel Johnson,
Boris Kopeliovich and Magno Machado for valuable discussion.
This work was supported by
the U.S.~Department of Energy at Los Alamos
National Laboratory under Contract No.~W-7405-ENG-38. 

\def\appendix{\par
 \setcounter{section}{0}
\setcounter{subsection}{0}
 \def\thesection{Appendix \Alph{section}}
\def\thesubsection{\Alph{section}.\arabic{subsection}}
\def\theequation{\Alph{section}.\arabic{equation}}
\setcounter{equation}{0}}

\appendix

\section{Derivation of the dipole formula for DY}\label{app:deriv}

In this appendix, we give a detailed derivation of the formula
that expresses the DY cross section in terms of the color dipole cross
section, Eq.~(\ref{eq:dylctotal}). This formula was first discovered in \cite{boris}.
Some details of the derivation in the semiclassical approach of \cite{semi}
are given in the appendix of \cite{bhq}. In this paper, we use the
framework of covariant Feynman perturbation theory.
The calculation can be carried out in a very 
similar way for DIS.

The differential cross section for dilepton production in the target rest frame, 
see Fig.~\ref{fig:dy}, 
is given by
\beqn\nonumber
d^8\sigma(qN\to ql^+l^-X)
\label{dsigma2}
&=&\sum_X\sum_{\lambda\lambda^\prime}
\epsilon^*_{\mu}(\lambda)\epsilon_\nu(\lambda^\prime)\overline{\cal M}^{\mu\nu}
\frac{d\alpha d^2q_\perp d^2p_{f\perp}}{(2\pi)^58(p_i^0)^2\alpha(1-\alpha)}\\
&\times &
\alpha_{em}
\epsilon_{\kappa}(\lambda)\epsilon^*_{\rho}(\lambda^\prime)
L^{\rho\kappa}\frac{dM^2d\Omega}{16\pi^2M^4},
\eeqn
where $p_i$ is the momentum of the projectile quark, $p_f$ is the momentum of
the final quark and $q$ is the momentum of the massive photon, 
$q^2=M^2>0$. 
The first line describes the bremsstrahlung of a massive photon from a quark
(as depicted in Fig.~\ref{fig:dy}),
and the second line the decay into the lepton pair $l^+l^-$ into solid angle $d\Omega$
(not shown in Fig.~\ref{fig:dy}). 
The electromagnetic coupling constant is denoted by
$\alpha_{em}=e^2/(4\pi)=1/137$. Furthermore, $\epsilon_\mu(\lambda)$ is
the polarization vector of the photon, $\lambda\in\{\pm1,0\}$.
We sum over all final states $X$ and
the photon polarizations $\lambda$, $\lambda^\prime$. 
The $\delta$-function for energy conservation is 
already integrated over. Furthermore,
 $\alpha$ is the energy fraction the photon takes
from its parent quark. It is assumed that only transverse momentum but no energy is
exchanged with the target.
Then, $\overline{\cal M}^{\mu\nu}$ is the absolute square of the
bremsstrahlungs-amplitude, before contraction with the polarization vector of
the photon. There are two contributions to this amplitude,
\beq
\overline{\cal M}^{\mu\nu}=
\frac{1}{2}\sum_{\sigma_f\sigma_i}
\frac{1}{N_c}\sum_{c_fc_i}
({\cal M}^{\mu}_s+{\cal M}^{\mu}_u)
({\cal M}^{*\nu}_s+{\cal M}^{*\nu}_u).
\eeq
The $s$-channel amplitude ${\cal M}^{\mu}_s$
describes the process where the photon is radiated
after the quark has scattered off the target, Fig.~\ref{fig:dy} (left), 
and the $u$-channel amplitude
${\cal M}^{\mu}_u$
describes the process where the photon is radiated before the quark scatters
off the proton, Fig.~\ref{fig:dy} (right). 
The bar indicates that the 
matrix element is summed over helicity $\sigma_f$ and color $c_f$
of the final quark and averaged over helicity $\sigma_i$ and color $c_i$
of the initial quark. ($N_c=3$ is the number of colors.) 
Finally, $p_{l^\pm}$ are the dilepton momenta and
\beq
L^{\mu\nu}=4(p^\mu_{l^+}p^\nu_{l^-}+p^\nu_{l^+}p^\mu_{l^-}
-g^{\mu\nu}p_{l^+}p_{l^-})
\eeq
is the leptonic tensor, summed over helicities. 

Different terms in the sum over photon helicities lead to different angular
distributions of the lepton pair. In this paper,
we integrate over the solid angle $d\Omega=d\phi d(\cos\theta)$, 
and obtain for the differential DY cross
section
\beqn\label{form}\nonumber
\frac{d^4\sigma(qN\to l^+l^-X)}{d{\rm ln}\alpha dM^2 d^2q_\perp }
&=&\frac{\alpha_{em}}{3\pi M^2}
\left\{\frac{d^3\sigma_{T}(qN\to\gamma^*X)}{d\ln\!\alpha\, d^2q_\perp}
+\frac{d^3\sigma_{L}(qN\to\gamma^*X)}{d\ln\!\alpha\, d^2q_\perp}\right\},
\eeqn
where the indices $T$ and $L$ stand for transverse and longitudinal photons,
respectively. We also integrated over the phase space of the final quark.
The factor $\alpha_{em}/(3\pi M^2)$ originates from the second
line of Eq.~(\ref{dsigma2}) and describes the decay of the $\gamma^*$ 
into the lepton pair. 
The interesting physics resides in the quantities 
$d^3\sigma_{T,L}/d\ln\!\alpha\, d^2q_\perp$, which are
given by
\beqn\label{eq:gammat}
\frac{d^3\sigma_{T}(qN\to\gamma^*X)}{d\ln\!\alpha\, d^2q_\perp}
&=&\int d^2p_{f\perp}\sum_X\sum_{\lambda\in\{\pm 1\}}
\frac{\epsilon^*_{\mu}(\lambda)\epsilon_\nu(\lambda)\overline{\cal M}^{\mu\nu}}
{(2\pi)^58(p_i^0)^2(1-\alpha)}\quad,\\
\label{eq:gammal}
\frac{d^3\sigma_{T}(qN\to\gamma^*X)}{d\ln\!\alpha\, d^2q_\perp}
&=&\int d^2p_{f\perp}\sum_X
\frac{\epsilon^*_{\mu}(\lambda=0)\epsilon_\nu(\lambda=0)\overline{\cal M}^{\mu\nu}}
{(2\pi)^58(p_i^0)^2(1-\alpha)}\quad.
\eeqn 
Terms with $\lambda\neq\lambda^\prime$ in Eq.~(\ref{dsigma2})
vanish after integration over
the azimuthal angle $\phi$.

We will now express the $d^3\sigma_{T,L}/d\ln\alpha d^2q_\perp$ in terms of
the $q\to\gamma^*q$ light-cone(LC) wavefunctions and the dipole cross section.  
Consider the $s$-channel graph in Fig.~\ref{fig:dy} (left)
where the photon is radiated after the
projectile quark has scattered off the target.
The propagator of the quark in the intermediate state can be written as
\beq
\frac{p_f\!\!\!\!\!/+q\!\!\!/+m_f}{(p_f+q)^2-m_f^2}=\sum_\sigma
\frac{u_\sigma(p_f+q)\bar u_\sigma(p_f+q)}{(p_f+q)^2-m_f^2}
-\frac{\gamma^+}{2(p_f^++q^+)}.
\eeq
The $u_\sigma(p)$ are
Dirac-spinors, $(p\!\!\!/-m_f)u_\sigma(p)=0$, for on-shell momentum $p$ and
helicity $\sigma$. The
$\gamma^+$-term arises, because the intermediate quark is off-shell. 
In the high
energy approximation which we employ, this term is dropped\footnote{As already
pointed out in \cite{bhq}, this approximation is equivalent to neglecting
the instantaneous vertices in a Hamiltonian light-cone approach to QCD
\cite{pauli}, and violates gauge invariance. The color dipole approach is gauge 
invariant only in leading log($x$) approximation.}. This is the crucial 
step that allows one to write the amplitude as a product of an LC wavefunction
and a quark scattering amplitude.

The $s$-channel amplitude then turns out to be
\beq
{\rm i}{\cal M}_s^\mu=
e\,\sum_\sigma
\frac{\bar u_{\sigma_f}(p_f)\gamma^{\mu}u_\sigma(p_f+q)}{(p_f+q)^2-m_f^2}\,
t_{q,\sigma\sigma_i}((p^0_f+q^0),\vec k_\perp).
\eeq
For simplicity, we have set the flavor charge to unity. Here,
$t_{q,\sigma\sigma_i}((p^0_f+q^0),\vec k_\perp)$ is 
the amplitude for scattering 
a quark off a nucleon in the rest frame of the nucleon, 
\beq
t_{q,\sigma\sigma_i}((p^0_f+q^0),\vec k_\perp)=
\bar u_\sigma(p_f+q)\,\gamma^0\, V_q(\vec k_\perp)\, u_{\sigma_i}(p_i)
\approx 2p_i^0\delta_{\sigma,\sigma_i}V_q(\vec k_\perp).
\eeq
Note that $t$ and $V$ are matrices in color space, we suppressed the color
indices.
Our spinors are normalized to 
$u_\sigma^\dagger(p)u_{\sigma^\prime}(p)=2p^0\delta_{\sigma,\sigma^\prime}$.
The function $V_q(\vec k_\perp)$ will be completely absorbed into the dipole cross section,
so that we will never need to specify it. Therefore, the dipole formulation also accounts
for some higher order and nonperturbative effects, which are difficult or even impossible
to account for in the standard parton model.
We write $V_q$ only as function of the exchanged transverse momentum
$\vec k_\perp=\vec p_{f\perp}+\vec q_\perp-\vec p_{i\perp}$, because
the longitudinal momentum of the projectile quark cannot be changed
significantly at
high energies. Note that $V_q$ also depends on energy, even though we do not write
this dependence explicitly.
The Kronecker-$\delta_{\sigma,\sigma_i}$ means that the
helicity of the quark is not changed by scattering off the proton.

In impact parameter space, the amplitude reads
\beqn\label{impact}
\widetilde{\cal M}^\mu_s(\vec b,\vec\rho)&=&
\int\frac{d^2l_{\perp} d^2k_\perp}{(2\pi)^4}
{\rm e}^{-{\rm i}\vec l_{\perp}\cdot\alpha\vec\rho-{\rm i}\vec k_\perp\cdot\vec b}
{\cal M}^\mu_s(\vec l_{\perp},\vec k_\perp)\\
&=&-\imag {\sqrt{4\pi}}\frac{\sqrt{1-\alpha}}{\alpha^2}
\Psi^{\mu}_{\gamma^* q}(\alpha,\vec\rho)
2p_i^0\widetilde V_{q}(\vec b),
\eeqn
where $\vec l_\perp=\vec p_{f\perp}-(1-\alpha)\vec q_\perp/\alpha$ 
is the transverse momentum of the final quark in a frame where the $z$-axis is
parallel to the photon momentum. The conjugate variable to $\vec l_\perp$, namely $\alpha\vec \rho$,
is then the transverse distance between initial and final quark in that frame. This quantity
will become the distance variable the dipole cross section. Furthermore,
\beq
\widetilde V_{q}(\vec b)=
\int\frac{d^2k_\perp}{(2\pi)^2}
{\rm e}^{-{\rm i}\vec k_\perp\cdot\vec b}V_q(\vec k_\perp).
\eeq
The LC wavefunctions in impact parameter space are related to the quark-photon
vertex and the propagator by Fourier transformation,
\beq\label{lcwavefunction}
\Psi^{\mu}_{\gamma^* q}(\alpha,\vec\rho)={\alpha^3}\sqrt{1-\alpha}
\int\frac{d^2l_{\perp}}{(2\pi)^2}{\rm e}^{-{\rm i}\vec l_{\perp}\cdot\alpha\vec\rho}
\sqrt{\alpha_{em}}\,
\frac{\bar u_{\sigma_f}(p_f)\gamma^{\mu}u_{\sigma_i}(p_f+q)}
{\alpha^2l_{\perp}^2+\eta^2},
\eeq
where $\eta^2=(1-\alpha)M^2+\alpha^2m_f^2$. In our numerical calculations in this 
paper, we set the quark mass $m_f=0$. See \cite{krt4} for its influence.

The $u$-channel graph can be written in the same way as (\ref{impact}), but
with a quark scattering amplitude at a shifted impact parameter. This can be
seen from the propagators.
While the propagator for the $s$-channel graph yields
\beq
\frac{1}{(p_f+q)^2-m_f^2}=\frac{\alpha(1-\alpha)}{\alpha^2l_{\perp}^2+\eta^2},
\eeq
one finds for the $u$-channel graph the combination 
$\vec l_{\perp}+\vec k_\perp$ instead of $l_{\perp}$,
\beq
\frac{1}{(p_i-q)^2-m_f^2}=-\frac{\alpha}
{\alpha^2(\vec l_{\perp}+\vec k_\perp)^2+\eta^2}.
\eeq
One then obtains for the $u$-channel amplitude in impact parameter space
\beqn
\widetilde{\cal M}^\mu_u(\vec b,\vec\rho)
&=&\imag\frac{\sqrt{4\pi}}{\alpha^2\sqrt{1-\alpha}}
\Psi^{\mu}_{\gamma^* q}(\alpha,\vec\rho)
2p_f^0\widetilde V_q(\vec b+\alpha\vec\rho)\\
&=&\imag\sqrt{4\pi}\frac{\sqrt{1-\alpha}}{\alpha^2}
\Psi^{\mu}_{\gamma^* q}(\alpha,\vec\rho)
2p_i^0\widetilde V_q(\vec b+\alpha\vec\rho).
\eeqn
and with Eqs.~(\ref{eq:gammat}) and 
(\ref{eq:gammal}) for the photon production cross sections
\beqn
\label{eq:gammat2}
\frac{d^3\sigma_{T,L}(qN\to\gamma^*X)}{d\ln\!\alpha\, d^2q_\perp}
&=&\int d^2p_{f\perp}
\frac{1}{(2\pi)^4}\int d^2b_1\int d^2b_2\int d^2\rho_1\int d^2\rho_2\\
\nonumber&\times&
\,{\rm e}^{{\rm i}\vec l_{\perp}\cdot\alpha(\vec\rho_1-\vec\rho_2)
+{\rm i}\vec k_{\perp}\cdot(\vec b_1-\vec b_2)}
\sum_{\lambda}\frac{1}{2}\sum_{\sigma_f\sigma_i}
\Psi^{\lambda}_{\gamma^* q}(\alpha,\vec\rho_1)
\Psi^{\lambda*}_{\gamma^* q}(\alpha,\vec\rho_2)\\
\nonumber&\times&
\sum_X
\frac{1}{N_c}\sum_{c_fc_i}
\left\{\widetilde V_q(\vec b_1)
-\widetilde V_q(\vec b_1+\alpha\vec \rho_1)\right\}
\left\{\widetilde V^\dagger_q(\vec b_2)
-\widetilde V^\dagger_q(\vec b_2+\alpha\vec \rho_2)\right\},
\eeqn 
where $\vec l_\perp=\vec p_{f\perp}-(1-\alpha)\,\vec q_\perp/\alpha$ and
$\vec k_\perp=\vec p_{f\perp}+\vec q_\perp-\vec p_{i\perp}$.
In the transverse case ($T$), the sum over $\lambda$ includes both transverse
polarizations, $\lambda=\pm 1$, while in the longitudinal case ($L$)
only one term,
$\lambda=0$ appears in the sum.

The LC wavefunctions for radiation of transverse and longitudinal photons
are readily calculated by contracting 
Eq.~(\ref{lcwavefunction}) with the photon polarization vector
\cite{boris,bhq,kst1},
\beqn\nonumber\label{dylct}
\Psi^{T}_{\gamma^* q}(\alpha,\vec\rho_1)
\Psi^{T*}_{\gamma^* q}(\alpha,\vec\rho_2)&=&
\sum_{\lambda=\pm 1}\frac{1}{2}\sum_{\sigma_f\sigma_i}
\epsilon^*_\mu(\lambda)\Psi^{\mu}_{\gamma^* q}(\alpha,\vec\rho_1)
\epsilon_\mu(\lambda)\Psi^{\mu*}_{\gamma^* q}(\alpha,\vec\rho_2)\\
\nonumber
&=& \frac{\alpha_{em}}{2\pi^2}\Bigg\{
     m_f^2 \alpha^4 {\rm K}_0\left(\eta\rho_1\right)
     {\rm K}_0\left(\eta\rho_2\right)\\
   &+& \left[1+\left(1-\alpha\right)^2\right]\eta^2
   \frac{\vec\rho_1\cdot\vec\rho_2}{\rho_1\rho_2}
     {\rm K}_1\left(\eta\rho_1\right)
     {\rm K}_1\left(\eta\rho_2\right)\Bigg\},\\
\label{dylcl}\nonumber
\Psi^{L}_{\gamma^* q}(\alpha,\vec\rho_1)
\Psi^{L*}_{\gamma^* q}(\alpha,\vec\rho_2)&=&
\frac{1}{2}\sum_{\sigma_f\sigma_i} 
\epsilon^*_\mu(\lambda=0)\Psi^{\lambda=0}_{\gamma^* q}(\alpha,\vec\rho_1)
\epsilon_\mu(\lambda=0)\Psi^{*\lambda=0}_{\gamma^* q}(\alpha,\vec\rho_2)\\
&=& \frac{\alpha_{em}}{\pi^2}M^2 \left(1-\alpha\right)^2
  {\rm K}_0\left(\eta\rho_1\right)
     {\rm K}_0\left(\eta\rho_2\right). 
\eeqn
Often, one needs only the sum over all polarization states of the photon.
We denote the corresponding quantity by
\beq\label{eq:lcwfsum}
\Psi_{\gamma^* q}(\alpha,\vec\rho_1)
\Psi^{*}_{\gamma^* q}(\alpha,\vec\rho_2)=
\Psi^{T}_{\gamma^* q}(\alpha,\vec\rho_1)
\Psi^{T*}_{\gamma^* q}(\alpha,\vec\rho_2)+
\Psi^{L}_{\gamma^* q}(\alpha,\vec\rho_1)
\Psi^{L*}_{\gamma^* q}(\alpha,\vec\rho_2).
\eeq

Integrating (\ref{eq:gammat2}) over the transverse momentum
$q_\perp$ of the photon yields
\beq\label{dylctotal}\nonumber
\frac{d\sigma_{T,L}(qN\to \gamma^*X)}{d{\rm ln}\alpha}=
\int d^2\rho\left|\Psi^{T,L}_{\gamma^* q}(\alpha,\vec\rho)\right|^2
\sigma^N_{q\bar q}(\alpha\rho),
\eeq
with the dipole cross section
\beq\label{eq:dcs}
\sigma_{q\bar q}(\alpha\rho)=
\sum_X\frac{1}{N_c}\sum_{c_fc_i}\int d^2b\left|\widetilde V_q(\vec b)
-\widetilde V_{q}(\vec b+\alpha\vec \rho)\right|^2.
\eeq
If one assumes that the quark interacts via one-gluon exchange 
with the target and expands Eq.~(\ref{eq:dcs}) to first order in
$(\alpha\rho)^2$, keeping only terms of order $\alpha_s$, 
then the dipole cross section is proportional 
to the target gluon density 

We can also express the transverse momentum distribution of DY pairs
in terms of the dipole cross section \cite{kst1}. 
The differential cross section is given by the 
Fourier integral
\beqn\nonumber\label{dylcdiff}
\frac{d^3\sigma_{T,L}(qN\to \gamma^*X)}{d\ln\alpha d^2q_\perp}
&=&\frac{1}{(2\pi)^2}
\int d^2\rho_1d^2\rho_2\, 
{\rm e}^{{\rm i}\vec q_\perp\cdot(\vec\rho_1-\vec\rho_2)}
\Psi^{*T,L}_{\gamma^* q}(\alpha,\vec\rho_1)
\Psi^{T,L}_{\gamma^* q}(\alpha,\vec\rho_2)\\
&\times&
\frac{1}{2}
\left\{\sigma^N_{q\bar q}(\alpha\rho_1)
+\sigma^N_{q\bar q}(\alpha\rho_2)
-\sigma^N_{q\bar q}(\alpha(\vec\rho_1-\vec\rho_2))\right\}.
\eeqn
To derive this expression, one performs the integration over 
$\vec p_{f\perp}$ in Eq.~(\ref{eq:gammat2})
and observes that Eq.~(\ref{eq:gammat2}) has a real value. 
This allows one to symmetrize the
integrand with respect to $\vec\rho_1$ and $\vec\rho_2$. The functions $V_q$
combine then to the dipole cross sections in the second line of 
Eq.~(\ref{dylcdiff}).
After integrating Eq.~(\ref{dylcdiff}) over the transverse momentum
$q_\perp$ of the photon, one obviously recovers
Eq.~(\ref{dylctotal}).

\end{document}